\documentclass[useAMS,usenatbib]{mn2e}

\usepackage{graphicx}
\usepackage{amsmath}
\usepackage{amssymb}

\usepackage{lineno,xcolor}
\newcommand{\poubelle}[1]{}

\title[Supernova remnants in very--high--energy gamma--rays]{Supernova remnants in the very--high--energy gamma ray domain: the role of the Cherenkov Telescope Array}
\author[P. Cristofari et al.]
{P. Cristofari$^{1}$\thanks{E-mail:~pc2781@columbia.edu},
S. Gabici$^{2}$, T. B. Humensky$^{3}$, M. Santander$^{4}$, R. Terrier$^{2}$ \and 
E. Parizot$^{2}$, S. Casanova$^{5,6}$
\\
$^{1}$Columbia University, 10027 New York, USA\\
$^{2}$APC, AstroParticule et Cosmologie, Universit\'e Paris Diderot, CNRS/IN2P3, CEA/Irfu, Observatoire de Paris, Sorbonne Paris Cit\'e,\\75205 Paris Cedex 13, France \\
$^{3}$Physics Department, Columbia University, 10027 New York, USA\\
$^{4}$Barnard College, Columbia University, 10027 New York, USA\\
$^{5}$Max Planck Institute for Nuclear Physics, D-69117 Heidelberg, Germany\\
$^{6}$Institute for Nuclear Physics, 31-342 Krakow, Poland
}
\begin{document}

\date{}

\pagerange{\pageref{firstpage}--\pageref{lastpage}} \pubyear{}

\maketitle

\label{firstpage}

\begin{abstract}
Supernova remnants are often presented as the most probable sources of Galactic cosmic rays. This idea is supported by the accumulation of evidence that particle acceleration is happening at supernova remnant shocks. Observations in the TeV range have especially contributed to increase the understanding of the mechanisms, but many aspects of the particle acceleration at supernova remnant shocks are still debated. 
The Cherenkov Telescope Array is expected to lead to the detection of many new supernova remnants in the TeV and multi--TeV range. In addition to the individual study of each, the study of these objects as a population can help constrain the parameters describing the acceleration of particles and increase our understanding of the mechanisms involved.

 %up to PeV energies (1 PeV=$10^{15}$ eV). 
\end{abstract}

\begin{keywords}
cosmic rays -- gamma rays -- ISM: supernova remnants.
\end{keywords}

\section{Introduction}
\label{sec:introduction}

Galactic cosmic rays (CRs) are relativistic particles (mainly protons) that fill the entire Galaxy and reach the Earth as an isotropic flux of particles. They are believed to be accelerated at supernova remnant (SNR) shocks via diffusive shock acceleration \citep[e.g.][]{drury83,bellCTA}. This idea is very popular but still needs to be conclusively proven. In order to be the sources of CRs, SNRs must convert $\approx 10\%$ of the supernova explosion energy into relativistic particles, with a spectrum somewhat steeper than $\propto E^{-2}$. Moreover, SNRs must be able to accelerate protons up to the energy of the CR "knee", located at a particle energy in the range $\approx 1- 4 \times 10^{15}$~eV~as suggested from experimental results~\citep{antoni2005,bartoli2015}. Finally, the observed chemical composition and the high degree of isotropy of CRs have to be explained as well~\citep[see e.g.][for reviews]{hillas,drury2012}. 

The success of the SNR paradigm for the origin of CRs resides in the fact that, within this framework, all the above mentioned conditions can be satisfied reasonably well.
Recent studies of non--linear diffusive shock acceleration seem to indicate that SNRs can indeed accelerate CRs with roughly the required efficiency, energy spectrum, and chemical composition \citep[see e.g.][]{seo}. As a result of the acceleration of CRs, the magnetic field is expected to be strongly amplified at SNR shocks, up to levels that would allow the acceleration of protons up to the energy of the knee \citep{bell04}. This scenario is supported by the fact that magnetic field strengths in the range from hundreds of microGauss to few milliGauss have been inferred from X--ray observations of a number of SNRs \citep[see e.g.][for a review]{jaccoreview}. The high level of isotropy of the arrival direction of CRs at Earth can also be reproduced under certain assumptions on the CR propagation in the Galaxy \citep{ptuskinanisotropies,blasiamato}. Recently, fine structures have been discovered in the CR spectrum \citep{pamela,cream} that impose additional constraints on the CR acceleration mechanism at SNR shocks \cite[see e.g.][]{ptuskinbreak}.   

One of the most promising ways to prove or falsify the scenario described above is to search for the radiation produced by CRs accelerated at SNR shocks.
Unambiguous evidence for the acceleration of particles at SNR shocks comes from the observations of individual SNRs in the radio, X-ray, and gamma--ray energy domains \cite[see][for a review]{eveline}. Synchrotron emission has been detected from many SNRs in both the radio \citep[e.g.][]{dubner} and the X-ray bands \cite[e.g.][]{jaccoreview}, and this tells us that SNRs can accelerate electrons at least up to the multi--TeV energy range \citep[e.g.][]{koyama}. On the other hand, the gamma--ray emission detected from SNRs can be interpreted either as the result of hadronic interactions of accelerated protons with the ambient gas, or as inverse Compton scattering of electrons on photons of the cosmic microwave background \citep[e.g.][]{felixreview}. Thus, in general it is hard to draw firm conclusions about the acceleration of CR protons from gamma--ray observations only. Recently, conclusive evidence has been obtained for the hadronic nature of the gamma--ray emission detected by Fermi and AGILE from several older SNRs interacting with molecular clouds \citep{fermipion,agilepion,jogler2016}, demonstrating that such objects are capable of accelerating CR protons in the GeV energy domain.

The detection of a handful of young (i.e. age up to a few thousands years) SNRs in TeV gamma--rays \citep{frankreview} was long sought and considered a crucial test for the SNR hypothesis for the origin of CRs \citep{dav,nt}. However, as explained above, such detections do not constitute a final proof for the SNR hypothesis, because it is hard to discriminate between gamma--rays produced in hadronic processes or as the result of inverse Compton scattering of electrons off soft ambient photons. Hence, the need for further tests to clarify in an unambiguous way whether SNRs are the sources of Galactic CRs.

%This is because if SNRs are the sources of CRs they must accelerate CR protons with an efficiency of $\approx 10\%$. The gamma--ray emission expected as the result of the interactions of such protons with the ambient gas (with typical density of $\approx 0.1...1$~cm$^{-3}$) compressed at the SNR shock is at a level which is detectable by current gamma--ray instruments.
%However, as stressed above, it is hard to discriminate between gamma--rays produced in hadronic processes or as the result of inverse Compton scattering of electrons off soft ambient photons. 
%Thus, the detection of young SNRs in TeV gamma rays supports the SNR hypothesis but does not constitute a proof for it. 

In \citet{cristofari1}, hereafter C13, we proposed a novel test for the SNR hypothesis based on the study of Galactic SNRs as a population. To do so, we performed Monte Carlo simulations to estimate the number of SNRs that one would expect to detect with Cherenkov telescopes of current generation, under the assumption that SNRs indeed are the main sources of Galactic CRs. We then compared our results with the number of actual detections of SNRs within the survey of the Galactic plane performed by the H.E.S.S. collaboration \cite[see e.g.][]{scan3}. The agreement found between our predictions and data brings further support to the SNR hypothesis. Moreover, physical constraints to the SNR scenario could be extracted. Most notably, we found that if the SNR hypothesis is correct, then the spectrum of accelerated particles at SNR socks must be steeper than $\approx E^{-2}$ and harder than $\approx E^{-4}$, because an $E^{-2}$ spectrum would result in a number of detections of SNRs much larger than the actual one, and $E^{-5}$ would result in too few detections. Additional constraints include the fact that a significant fraction (roughly $\approx 60 \%$ or more, according to model parameters) of the SNRs detected in the H.E.S.S. survey are expected to be characterized by hadronic emission, and the finding that thermonuclear supernovae (type Ia) are expected to account for a large fraction ($\approx 60 - 80 \%$) of the detections. For a more detailed discussion of these findings we refer the reader to C13.

The results summarized above indicate that the approach we proposed in C13 is powerful, despite the fact that only very few SNRs have been firmly identified within the H.E.S.S. survey of the Galactic plane. With time, new SNR shells~\citep{HESS_GPS_2016} are being detected, and the increased statistics can be confronted with our results to test more accurately and constrain even further the SNR scenario.
These larger statistics are expected to be brought by instruments of the next generation such as the Cherenkov Telescope Array (CTA). 

Thus, the main objective of this paper is to evaluate the impact that next--generation instruments such as CTA, operating in the TeV and multi--TeV domain, will have on these studies. After reviewing briefly our method in Sec.~\ref{sec:montecarlo}, we will make predictions on the number of SNRs detectable by future gamma--ray telescopes. These predictions will be based on the assumption that SNRs are the main sources of CRs. The consistency between our previous results published in C13 and the data currently available from the H.E.S.S. Galactic Plane Survey (GPS)~\citep{HESS_GPS_2016} argues for the robustness of the predictions presented in this paper. In Sec.~\ref{sec:CTA} we will consider the case of the GPS of the Cherenkov Telescope Array, whose sensitivity in the TeV energy range is expected to be significantly improved compared to current Cherenkov telescopes~\citep{hinton2013}. We describe how Monte Carlo simulations confronted with future observations of CTA shall help constrain the parameters involved in the acceleration of VHE particles at SNR shocks, and thus provide a better understanding of the origin of Galactic cosmic rays. 

%The opening of the multi--TeV ow in the electromagnetic spectrum will have a dramatic impact in the search for the sources of CRs. This is because in this energy domain the efficiency of inverse Compton scattering is strongly reduced, due to the Klein--Nishina suppression of the interaction cross section \citep{gould}. Thus, the detection of a SNR exhibiting a gamma--ray spectrum which extends unattenuated until several tens or hundreds of TeV would imply that the nature of the gamma--ray emission is hadronic. Moreover, since the parent CR protons have particle energies a factor of $\approx$~10 larger than those of the gamma--rays, such a detection would also prove that the SNR is currently accelerating particles up to energies of hundreds of TeVs or several PeVs. In other words, we would have the proof that the SNR is acting as a PeVatron and capable of accelerating particles up to the energy of the knee. Due to the importance of this issue, we will devote Sec.~ref{} to investigate the capability of gamma--ray instruments of next generation in detecting and identifying PeVatrons. 

\section{Cosmic--ray acceleration and gamma--ray production in Galactic supernova remnants}
\label{sec:montecarlo}

In this section, we briefly describe the Monte Carlo approach used to simulate the time and position of explosion of supernovae within the Galaxy. This procedure is then coupled with a model which describes the acceleration of particles at SNRs and the related production of gamma rays. In this way, the number of SNRs detectable by a given gamma--ray instrument can be computed. This approach was presented in~C13, where a detailed description of the model can be found. 
%In the following we briefly summarize it, highlighting the main differences and improvements added in this new paper. 
%This procedure will be used in Section~\ref{sec:CTA} to predict the number of SNRs that telescopes of the next generation, such as the Cherenkov Telescope Array, are expected to detect. 

\subsection{Time and spatial distribution of supernovae in the Galaxy}

The time of explosion of all Galactic supernovae is simulated assuming that the supernovae explosion rate is constant in time and equal to $\nu_{\rm SN} = 3 /$century~\citep[see e.g.][and references therein]{li2011}. In fact, this value is quite uncertain within the range extending from about one to a few supernova explosions per century. However, as discussed in C13, results are quite insensitive to the actual choice of this parameter, if the value of the total CR luminosity of the Galaxy is fixed. 
A type is assigned to each supernova. Four types of supernova are considered: Ia, IIP, Ib/c and IIb.
 For each, typical values are assumed for the total explosion energy ${\cal E}$, the mass of the ejecta $M_{\rm ej}$, the mass--loss rate $\dot{M}$ and the velocity of the wind $u_{\rm w}$~\cite[see e.g.][]{seo}.

 % Assuming a relative rate of 0.04 as proposed in~\citep{seo} does not affect significantly the results of this paper. 
%\begin{table}
%\centering
%\begin{tabular}{c c c c c c}
%\hline 
%Type & ${\cal E}_{51}$ & $M_{ej,\odot}$ & $\dot{M}_{-5}$ & $u_{w,6}$ & Rel. rate\\
%\hline 
%\hline
%Ia & 1 & 1.4  & -- & -- & 0.32\\
%IIP & 1 & 8 & 1 & 1  & 0.44 \\
%Ib/c & 1 & 2 & 1 & 1 & 0.22 \\
%IIb & 3 & 1 &  10 & 1 & 0.02 \\
%\hline
%\end{tabular}
%\caption{Supernova parameters adopted in the simulation: supernova type (column 1), explosion energy in units of $10^{51}$~erg (column 2), mass of ejecta in solar masses (column 3), the wind mass loss rate in $M_{\odot}$/yr (column 4), the wind speed in units of 10~km/s (column 5), and the relative explosion rate (column 6). Values from \citet{seo}.}
%\label{tab:types}
%\end{table}
% 
 
Once the time of the explosion and the type of a supernova is drawn, a location within the Galaxy is assigned. The description proposed by~\citet{case} has been widely used in the literature, but was criticized by~\citet{green2015}.
Here the distribution of core--collapse supernovae is taken following the description of~\citet{faucher}, assuming that SNRs in the Galaxy follows the radial distribution of pulsars, as presented by~\citet{PSR,lorimer}. The surface density at the galactocentric radius $r$ is: 
\begin{equation}
\rho (r)=A \left( \frac{r+R_{\odot}}{R_{\odot}+R_1}\right)^{a} \text{exp} \left[ - b \left(Ê\frac{r-R_{\odot}}{R_{\odot} +R_1} \right) \right]
\label{eq:radialdistribution}
\end{equation}
where $R_{\odot} = 8.5 ~\text{kpc}$ is the Sun's galactocentric distance, $R_1=0.55$ kpc, $a=1.64$ and $b=4.01$ are model parameters taken from \citet{lorimer},  and $A$ is a normalisation constant determined by imposing a total supernova rate in the Galaxy equal to $\nu_{SN} = 3$/century. 
Four spiral arms are considered, with each arm following a logarithmic spiral shape. The centroids of the arms are described analytically by equations of the form \citep{faucher}:
\begin{equation}
\label{eq:arms}
\theta (r ) = k ~ \text{ln}(r/r_0)+\theta_0
\end{equation}
where $r$ is the galactocentric distance and $\theta$ is the polar angle. The parameters $k$, $r_0$, and $\theta_0$ are listed in Table~\ref{tab:spiralarms}.

The spiral structure is realised by drawing first a galactocentric distance $r_{\rm raw}$ from Eq.~\ref{eq:radialdistribution} and then by choosing randomly an arm. The polar angle $\theta_{\rm raw}$ is then determined so that the supernova lies in the centroid of the arm. The actual position of the supernova is finally computed by applying a correction $r_{\rm corr}$ to the galactocentric distance drawn from a normal distribution centered at zero with standard deviation $0.07\; r_{\rm raw}$.
To avoid artificial features near the Galactic center, the distribution is blurred by applying a correction to $\theta_{\rm raw}$ also, of magnitude $\theta_{\rm corr} \text{exp}(-0.35 r_{\rm raw}\text{kpc})$, where $\theta_{\rm corr}$ is randomly chosen in the interval $[0, 2\pi]$ rad.

\begin{table}
\centering
\begin{tabular}{c c c c c c}
\hline 
 & Name & $k$ [rad] & $r_0$ [kpc] & $\theta_0$ [rad] \\
\hline 
\hline
1 & Norma & 4.25  & 3.48 & 1.57\\
2 & Carina--Sagitarius & 4.25  & 3.48 & 4.71\\
3 & Perseus & 4.89  & 4.90 & 4.09\\
4 & Crux--Scutum & 4.89  & 4.90 & 0.95\\

\hline
\end{tabular}
\caption{Spiral arms parameters from~\citet{faucher}, to be used in Eq.~\ref{eq:arms}.
}
\label{tab:spiralarms}
\end{table}

The spatial distribution of thermonuclear supernovae (type Ia) cannot be traced by the distribution of pulsars, since their progenitors are old stars of relatively low mass.
We adopted here the radial distribution of supernovae type Ia plotted in Fig.~10 of \citet{prantzos2011}, which has been derived from the studies by \citet{scannapieco2005}. From estimation of the thermonuclear SN rate in the Galaxy~\citep{scannapieco2005}, \citet{prantzos2011} have parametrized the SNIa radial profile in the Galaxy. In our work, we consider this parametrized profile and in the inner regions of the Galaxy where the Prantzos description is incomplete (0--2 kpc) we assume a flat distribution normalized using the relative SN rates given by~\citet{mannucci2006}. 

The altitude of the SNRs above (or below) the Galactic plane is determined by assuming that the vertical distribution of the SNRs follows that of the gas~\citep{HI,H2}. We use the vertical distribution of molecular hydrogen for core--collapse supernovae, and the vertical distribution of atomic Hydrogen for thermonuclear supernovae. This implies that the distribution of thermonuclear supernovae extends up to a height above the disk which is significantly larger than that of core--collapse supernovae. In the absence of a complete knowledge of the spatial distribution of supernovae of a given type, this assumption accounts for the fact that core--collapse supernovae are expected to explode in dense star--forming regions, while thermonuclear ones can also be found in low density regions.

\subsection{Evolution of SNRs}
 The four considered types of progenitors can be divided into two sub-types: thermonuclear supernovae (type Ia) and core collapse supernovae (all other types), for which the dynamical evolution of the SNR shock is described by different expressions. To determine the time evolution of the SNR shock radius and velocity, we rely on the approach described in~\citet{pz03,pz05}, where the CR contribution to the pressure behind the SNR shock is assumed to be significant. 
 
In the case of a thermonuclear supernova, the time evolution of the shock radius $R_{\rm sh}$ and the shock velocity $u_{\rm sh}$ in the ejecta--dominated phase are described by self--similar expressions\citep{chevalier,pz05}. 
%\begin{equation}
%\label{eq:ejIa}
%\begin{split}
%&R_{\rm sh} = 5.3 \; {\left(  \frac{{\cal E}_{51}^2 \; M_{\odot} }{n_{0} M_{\rm ej}}  \right)}^{1/7}  \; t_{\rm kyr}^{4/7} \; \text{pc}  \\
%&u_{\rm sh}= 3.0 \times 10^3 \; \left(  \frac{{\cal E}_{51}^2 \; M_{\odot} }{n_{0} M_{\text{ej}}}  \right)^{1/7}   \; t_{\text{kyr}}^{-3/7} \; \text{km s}^{-1}
%\end{split}
%\end{equation}
%where ${\cal E}_{51}$ is the supernova explosion energy in units of $10^{51}$~erg, $n_0$ is the ambient gas number density in cm$^{-3}$, $M_{\rm ej}$ is the mass ejected in the explosion in solar mass units, and $t_{\rm kyr}$ is the time after explosion expressed in kilo--years.
The SNR evolution during the adiabatic phase is computed using the expression given in~\citet{truelove,pz05}. 
% \begin{equation}
%\label{eq:rshocktype1sedov1}
%\begin{split}
%&R_{\rm sh} = 4.3  \left(\frac{ {\cal E}_{51}}{n_{0}} \right)^{1/5} t_{\rm kyr}^{2/5}   \left( 1 - \frac{0.06 ~ M_{{\rm ej},\odot}^{5/6}}{{\cal E}_{51}^{1/2} n_0^{1/3} t_{\rm kyr}} \right)^{2/5}   \; \text{pc}  \\
%&u_{\rm sh}= 1.7 \; 10^3  \left(\frac{ {\cal E}_{51}}{n_{0}} \right)^{1/5} t_{\rm kyr}^{-3/5} \; \left( 1 - \frac{0.06 ~ M_{\rm ej,\odot}^{5/6}}{{\cal E}_{51}^{1/2} n_0^{1/3} t_{\rm kyr}} \right)^{-\frac{3}{5}}  \text{km s}^{-1}
%\end{split}
%\end{equation}
%which connect smoothly with Equations~\ref{eq:ejIa} at a time $t_0 \sim 260 (M_{\rm ej,\odot}/1.4)^{5/6} {\cal E}_{51}^{-1/2} n_0^{-1/3}$~yr, and tend to the exact Sedov--Taylor solution \citep{sedov,taylor} for $t \gg t_0$.
We follow the SNR evolution until the shock enters the radiative phase, typical after a few $10^4$ years~\citep{cioffi}.
In the case of a core--collapse supernova, the shock propagates in the wind--blown bubble generated by the wind of the progenitor star. \citet{pz05} suggests to model the wind blown bubble as two regions: a dense red--supergiant wind and a tenuous hot bubble which has been inflated by the wind of the massive progenitor star in main sequence~\citep{weaver,longair}. A more detailed description of the environment around core--collapse supernovae can be found in~C13.
A description of the evolution of the SNR shock during the ejecta dominated phase and the adiabatic phase in this structured interstellar medium is given in~\citet{chevalier,pz05}. In the adiabatic phase, the SNR shock evolution can be obtained by adopting the thin--shell approximation~\citep[e.g.][]{ostriker,bisnovati}. 

Finally, a crucial parameter is the density of the ambient medium, which we derived from the surveys of atomic and molecular hydrogen presented in~\citet{HI,H2}. Typical values of the ambient density are found in the range $\approx 10^{-5}- 10$ cm$^{-3}$. 

\subsection{Particle Acceleration and Gamma Emission at SNR Shocks}
\label{sec:gammafromSNRS}
%In the previous Sections, a model for the time and spatial distribution of SNRs in the Galaxy and for the dynamical evolution of SNRs has been assumed. 
The gamma--ray emission related to the acceleration of particles from a given SNR shell is computed following the method adopted in~C13. 
The contributions to the gamma--ray emission of both nuclei and electrons accelerated at the SNR shocks, through neutral--pion decay and inverse Compton scattering, respectively, are taken into account.
The acceleration of particles at the shock is assumed to follow a power--law spectrum $f(R_{\rm sh}, p, t)= A(t) p^{-\alpha}$ where $\alpha$ is regarded in the following as a free parameter. In order to reproduce the slope of the CR spectrum at Earth, studies have suggested that values for alpha should be in the range $\alpha=4.4- 4.1$~\citep[e.g.][]{vladimirdrift,damiano,pz05,ohira,damianodamping,stefanoescape} coupled with studies of the propagation of CRs in the Galaxy \citep[see e.g.][and references therein]{strongreview}. 
By equating the CR pressure at the shock to a fraction $\xi_{CR} \approx 0.1$ of the shock ram pressure, an expression for the normalization $A(t)$ of the CR spectrum can be obtained. 

The maximum energy of the accelerated particles is determined by assuming that particles escape the acceleration site once their diffusion length equates a fraction $\zeta$ of the shock radius. Several studies suggest $\zeta \approx 0.05 - 0.1$~\citep[e.g.][and references therein]{zirakashviliptuskin08}, and we here take $\zeta = 0.1$ as a reference value.

%
%The maximum momentum $p_{\rm max}$ of the accelerated particles is determined here by the %most stringent of the two following relations:
%equality:
%\begin{equation}
%%t_{acc} = \frac{D(p_{max})}{u_{sh}^2} \approx t_{age} \\
%\label{eq:confinement}
%l_{\rm d} = \frac{D(p_{\rm max})}{u_{\rm sh}} \approx \zeta R_{\rm sh}
%\end{equation}
%where $D$ represents the momentum--dependent diffusion coefficient for CRs upstream of the shock.
% In this expression $p_{\rm max}$ is defined as the momentum for which the diffusion length of particles ahead of the shock $l_{\rm d}$ equates to some fraction $\zeta$ of the shock radius. 
%Particles with larger momentum are characterized by larger diffusion length and are assumed to escape the SNR. Studies of particle acceleration and escape from shocks suggest that $\zeta \approx 0.05 - 0.1$ \citep[e.g.][and references therein]{zirakashviliptuskin08}. In the following we will adopt the value $\zeta = 0.1$.

At the shock, the particle diffusion coefficient depends on the structure and strength of the magnetic field.  A high efficiency of the CR acceleration is expected along with a strongly amplified magnetic field and a reduced diffusion coefficient with respect to the typical values found in the interstellar medium. Several mechanisms have been proposed to explain the magnetic field amplification, such as the CR current driven instability~\citep{bell04}, or the resonant streaming instability~\citep{lagage}, or the Drury instability~\citep{luke}. 
In the following, we will work under the assumption that, in the presence of efficient magnetic field amplification, CR obey Bohm diffusion in the amplified field. 
The  expressions adopted for the magnetic field downstream of the shock $B_{\rm down}$ and for the diffusion coefficient are detailed in~C13. 
Assuming that a fraction $\xi_{\rm B}$=3.5\% of the ram pressure is converted into magnetic energy, we get $B_{\rm down} \propto ( u_{\rm sh}/ v_{\rm d})$ where $v_{\rm d}$ is treated as a parameter scaling as $v_{\rm d} \propto \xi_{\rm B}^{-1/2}$. Other works have proposed different descriptions of the evolution of magnetic field at the shock. From X--ray observations, it was for example proposed to describe $B_{\rm down} \propto (u_{\rm sh}/ v_{\rm d})^{3/2}$~\citep{bell04,vink2008,bykov2014}, with a fraction $\xi_{\rm B}$ increased within a factor of 2, thus decreasing $v_{\rm d}$. In this description,  the amplified magnetic field can reach higher values, but for a shorter time~\citep[see e.g.][and reference therein]{jaccoreview}. Adopting such a description does not affect substantially the results presented in this paper.

Electrons are accelerated at the shock at the same rate as protons. Their spectrum is, however, different, because they suffer synchrotron and inverse Compton losses. At low energies, losses can be neglected, so the acceleration of electrons and protons proceeds similarly: the same spectral shape is thus expected for both species, and a parameter $K_{\rm ep}$ is introduced to describe the ratio between the electron and proton spectra. Indications of values in the range $\approx 10^{-4} -10^{-2}$ have been obtained from spectral fits of individual SNRs~\citep[see e.g.][]{donRXJ,giovannitycho}.  For smaller values ($K_{\rm ep}\approx 10^{-5}$ or less),  the gamma--ray emission from electrons becomes negligible in comparison to the emission due to pion decay~C13. For this reason, in the following we will consider the range $10^{-5} -10^{-2}$.
After being accelerated, the electrons are advected downstream of the shock where they lose energy, mainly through synchrotron radiation. 

%, with a characteristic time given by:
%\begin{equation}
%\tau_{\text{sync}} \approx 4.5 \times 10^4 \left(  \frac{E^{\text{e}}}{\text{TeV}}\right)^{-1}  \left( \frac{B_{\text{shell}}}{20 \mu\text{G}}\right)^{-2}  \text {year}
%\label{eq:tausyn}
%\end{equation}
%where $E^{\rm e}$ is the electron energy and $B_{\rm shell}$ is the magnetic field strength inside the SNR shell. Note that $B_{\rm shell}$ is expected to be significantly smaller than $B_{\rm down}$ (see Eq.~\ref{eq:amplifiedB}) if the magnetic field is damped while it is advected downstream of the shock~\citep[see e.g.][]{pohl,atoyantycho}.
The magnetic field inside the shell $B_{\rm shell}$ is expected to be significantly smaller than $B_{\rm down}$ if the  magnetic field is damped while it is advected downstream of the shock~\citep[see e.g.][]{pohl,atoyantycho}.
 In the following we use $B_{\rm shell}$= 20 $\mu$G, and we refer the reader to~C13 for a more extended discussion of this issue. 
Because the energy loss time decreases with particle energy, an energy $E^{\text{e}}_{\text{break}}$ exists above which the loss time is shorter the SNR age. Above such energy, the electron spectrum is shaped by radiative losses and steepens by one power in energy with respect to the injection spectrum~\citep[see for example][and references therein]{giovannitycho}. 
The maximum energy of the electrons $E^{\rm e}_{\text{max}}$ accelerated at a shock can then be obtained by equating the acceleration rate at the shock to the synchrotron energy loss time. 
%In our calculations, we follow the approach described in~\citet{vannoni2009}.  In the case $u_{\rm sh}>> v_{\rm d}$, we get as an illustrative example:
%\begin{equation}
%E^{\rm e}_{\text{max}} \approx 7.3 \left( \frac{u_{\text{sh}}}{1000 \text{ km/s}}\right)  \left(\frac{B_{\text{down}}}{100 \; \mu\text{G}} \right)^{-1/2} \; \text{TeV}
%\label{eq:emaxe}
%\end{equation}

The description of the acceleration of particles inside a SNR can finally be derived by solving a transport equation:
\begin{equation}
\frac{\partial f}{\partial t} +u \nabla f - \nabla D \nabla f - \frac{p}{3} \nabla u \frac{\partial f}{\partial p} = 0 
\label{eq:transport}
\end{equation}
where D is the momentum--dependent diffusion coefficient for CRs and $p$ the momentum of particles. 
To describe the structure of the interior of the SNR, we follow the approach of~\citep{pz03,pz05}. 
Inside the SNR, the structure can be determined by assuming that the velocity is a linear function of the radius, and by solving the gas continuity equation~\citep[see e.g.][]{ostriker}. Assuming a CR efficiency in the range 0.05--0.1, we adopt a compatible compression factor $\sigma = 4$. Working under the assumption of such a CR efficiency ensures the validity of the self-similar solutions proposed by~\citet{ostriker}. 
The effect of the plasma thermal conduction in the interior of the SNR~\citep[see e.g.][]{shelton1999} are not taken into account, but are not expected to be dominant, as our results depend mainly on the evolution of the shock radius and velocity, and not on the details of the internal structure. 

The reader can refer to~C13 for more details. We can therefore finally compute the gamma--ray luminosity from a given SNR. This is done by calculating the hadronic component from proton--proton interactions, following the approach of~\citet{kelner2006}, and multiplying results by a factor 1.8 to take into account the nuclei heavier than hydrogen present in the CRs and in the ambient gas~\citep{mori2009}. The leptonic component from inverse Compton scattering of accelerated electrons on the cosmic microwave background is then added~\citep{gould}.

%To determine the internal structure of the SNR, the linear velocity approximation is adopted~\citep{ostriker},  with a shock compression factor assumed to be  $\sigma = 4$ in the absence of CR acceleration for a strong shock. 

\subsection{Comparison with available TeV data: the H.E.S.S. survey of the Galactic plane}
\label{sec:previouswork}
In a previous work (Cristofari et al. 2013), a Monte Carlo procedure analogous to the one presented in this article was used to provide a novel test of the SNR paradigm for CR origin. In particular, comparisons with the H.E.S.S data from the GPS available at the time were used to confront our predictions. The choice of the values of two free parameters was shown to impact dramatically our results, namely, the spectral slope of the particles accelerated at the SNR shock $\alpha$, and the electron-to-proton ratio $K_{\rm ep}$. Plausible ranges for these values are $\alpha = 4.1- 4.4$ and $K_{\rm ep} = 10^{-5} - 10^{-2}$ (see Cristofari et al. 2013 for a discussion). Here and in the following we consider three scenarios, labeled M1, M2, M3, and characterized by the values of $\alpha$ and $K_{\rm ep}$ listed in Tab.~\ref{tab:modelsCTA}. Clearly, the scenario M1 (M3) would result in the largest (lowest) number of expected detections of SNRs.

To our knowledge, the most updated results on the H.E.S.S. GPS have been presented in~\citet{donath2016}. To date, 78 very-high-energy gamma-ray sources have been detected in the GPS. 31 of them have been firmly identified with known astrophysical objects, while the rest still remain unidentified, or only tentatively identified. Amongst the 31 firm identifications, 8 are SNRs and 8 are composite SNR sources (i.e. it is not clear whether the emission is produced by the SNR or by the associated pulsar wind nebula). Thus, a fraction between 1/4 and 1/2 of the firmly identified sources are SNRs. If this fraction is representative of the entire sample of GPS sources, then one might estimate that 20 to 40 out of the 78 GPS sources might be SNRs. Therefore, the most conservative (generous) estimate of the number of SNRs detected in the H.E.S.S. GPS is of $\lesssim 10$ (few tens).

Using the procedure presented in the previous Sections, we computed the number of expected detections of SNRs in the H.E.S.S. GPS, which within the portion of the Galactic disk defined by the coordinate ranges $60^{\circ} < l < 260^{\circ}$, $| b | < 2.5^{\circ}$ is characterized by a roughly uniform sensitivity equal to 1.5 \% of the Crab flux above 1 TeV. For the two extreme scenarios M1 and M3 we predict $36^{+7}_{-6}$ and $3.2^{+2}_{-2}$ detections of SNRs, respectively, compatible with the range inferred from observations.
%These models accounts for exploration of the range $K_{\rm ep}=10^{-5}.. 10^{-2}$ and $\alpha=4.1 .. 4.4$. The effects of the function $p_{\rm max}$ in discussed further. 

\begin{table}
\centering
\begin{tabular}{c c c }
\hline 
Model & ~~~~$\alpha$~~~~ & ~~~~$K_{\rm ep}$~~~~ \\
\hline 
\hline
M1 & 4.1 & $10^{-2}$    \\
M2 & 4.4 & $10^{-2}$   \\
M3 & 4.4 & $10^{-5}$   \\
\hline
\end{tabular}
\caption{Values of the parameters adopted to compute the curves in Fig.~\ref{fig:LNLS1}, Fig.~\ref{fig:LNLS2}, Fig.~\ref{fig:confusion}  and~Table~\ref{tab:numbersCTA}. 
$\alpha$ is the slope of the spectrum of CRs accelerated at the shock, and $K_{\rm ep}$ is the electron--to--proton ratio.}
\label{tab:modelsCTA}
\end{table}

\section{Detection of SNR shells with CTA} 
\label{sec:CTA}

The Monte--Carlo procedure described above is used to simulate the typical population of SNRs expected to be accessible to the Cherenkov Telesecope array. In the following, all results presented have been obtained by averaging 1000 Monte Carlo realizations of the Galaxy. In fact, a few hundred realizations of the Galaxy are sufficient to produce the results of this article, we take 1000 as a round number.

One of the main scientific goals of the Cherenkov Telescope Array will be to survey areas of the sky to search for faint VHE gamma--ray sources. Most of the currently known VHE sources are located in the Galactic plane. Although the final performance of the array will depend on the exact number of telescopes deployed, it has been shown that CTA could be able to carry out a survey of the entire Galactic plane, directed towards the study of the  region $| l  | < 60^{\circ}$, $| b | < 2^{\circ}$, in $\approx$1/4 of the available observation time per year, with a uniform sensitivity down to $\approx 3$  mCrab above 1 TeV~\citep{dubus2013}. This corresponds to a flux of $\approx 6.9 \times 10^{14}$ cm$^{-2}$ s$^{-1}$.   The details of this hypothetical GPS are still a matter of discussion, but we can consider these values as a reference. 

\begin{figure}
\includegraphics[width=.5\textwidth]{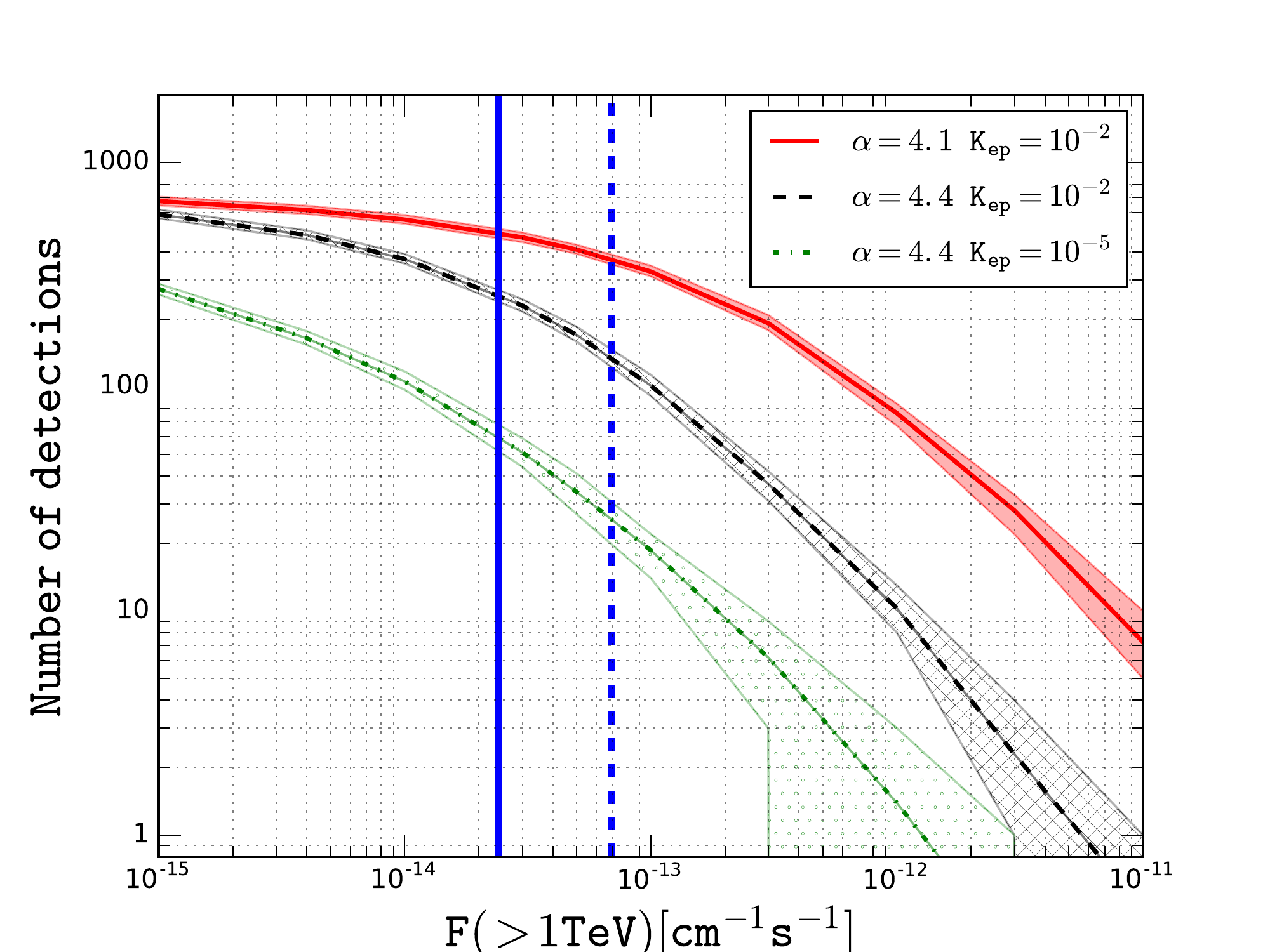}
\caption{SNRs in the entire Galaxy with integral gamma--ray flux above F($>$1 TeV). The red (solid) curve corresponds to model M1, the black (dashed) line corresponds to M2 and the green (dot--dashed) line to M3. In each case the +/- standard deviation is shown. The blue solid and dot--dashed vertical lines correspond to a sensitivity of 1 mCrab and 3 mCrab, respectively.}
\label{fig:LNLS1}
\end{figure}
\begin{figure}
\includegraphics[width=.5\textwidth]{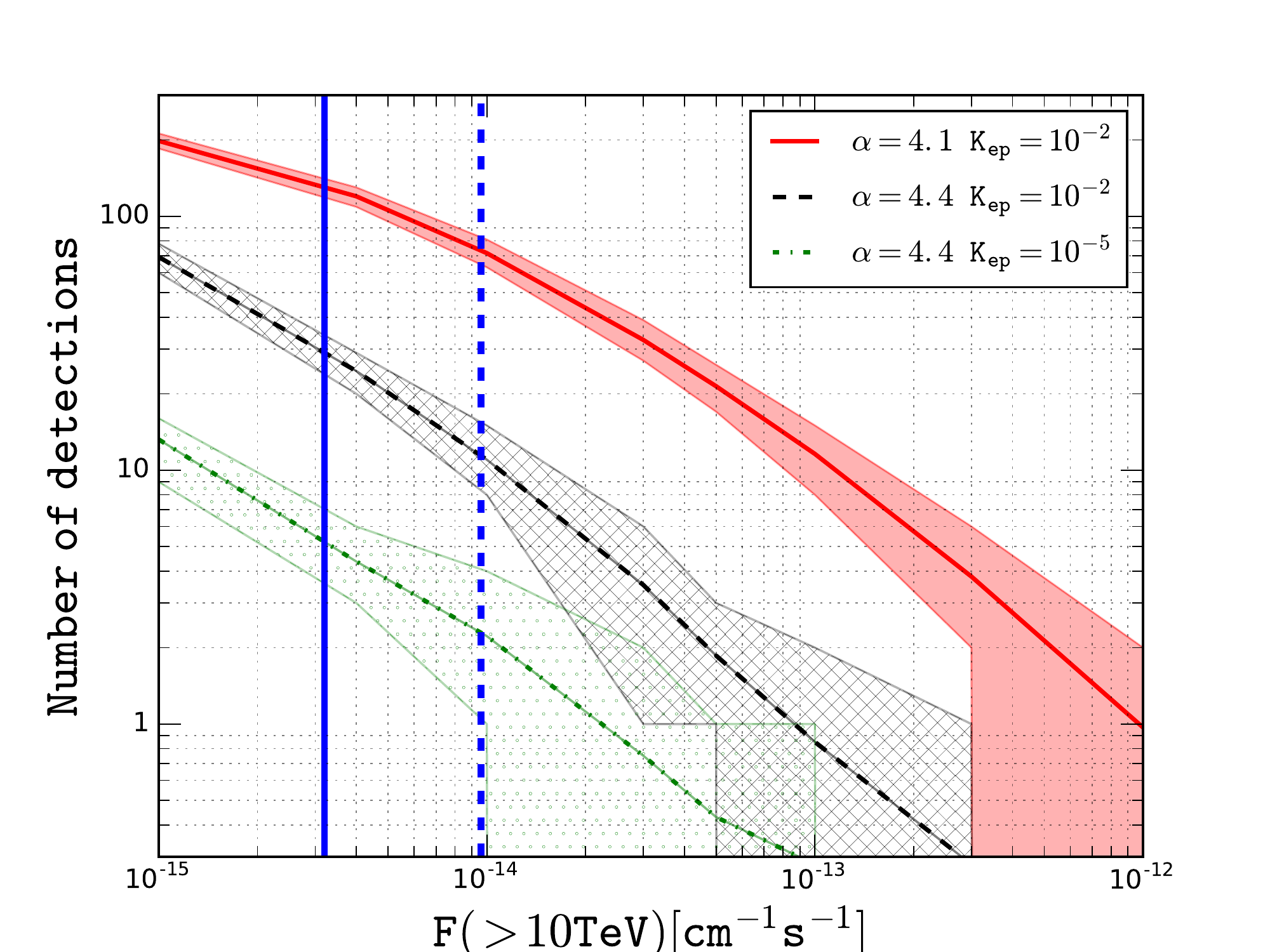}
\caption{SNRs in the entire Galaxy with integral gamma--ray flux above F($>$10 TeV). Curves as described in caption of Fig.~\ref{fig:LNLS1}. The blue solid and dot--dashed vertical lines correspond to a sensitivity of 10 mCrab and 30 mCrab, respectively.}
\label{fig:LNLS2}
\end{figure}

We start by computing the number of SNRs in the Galaxy with integral gamma--ray flux above a given value F($>$E). Results are shown in Fig.~\ref{fig:LNLS1} and Fig.~\ref{fig:LNLS2} for F($>1$ TeV) and F($>10$ TeV), respectively, for the three scenarios M1, M2, and M3 described above and summarized in Tab.~\ref{tab:modelsCTA}. 

The blue vertical lines correspond to the typical point--source sensitivity achieved by CTA (impact of extension is discussed below), namely $\approx 1$ mCrab above 1 TeV and $\approx 10$ mCrab above 10 TeV~\citep{dubus2013}. 
%As the sensitivity of the Cherenkov Telescope Array is expected to reach . above photon energies of 1 TeV, 10 TeV and 50 TeV respectively, the number of detectable SNRs is expected to grow., na
Fig.~\ref{fig:LNLS1} shows that, for photons of energies above 1 TeV, the models M1, M2 and M3 lead to an average number of SNRs potentially detectable by CTA in a pointed observation of 430, 220 and 48 respectively. At 10 TeV, these numbers are 120,  28 and 5.2. This suggests that under certain circumstances CTA might detect a non negligible fraction of the whole SNR population in the Galaxy, which can be estimated from the SN rate and a typical maximum age of SNRs of $\approx 10^{5}$ year, thus $\approx 3 \times 10^3$ potential SNRs. Many SNRs enter the radiative phase younger, suggesting an even larger fraction young SNRs should be detected. This might constitute the transition between a case--by--case study of gamma--ray bright SNRs to a real population study. 

The number of potentially detectable SNRs above 10 TeV is found between several tens or very few objects depending on the parameters adopted. This will have a crucial importance for the search for PeVatrons (SNRs accelerating PeV particles)~\citep[see e.g.][and reference therein]{gabici2016}.  
 %since the detection in this range is required for such objects

The numbers in Fig.~\ref{fig:LNLS1} and Fig.~\ref{fig:LNLS2} can be seen as an optimistic and ideal situation, and thus are upper limits for the numbers of detections.  
This is because several effects can be expected to reduce the number of detections. First of all, it has been described that the sensitivities considered above (represented by the vertical blue lines) will not be achieved in the entire Galactic plane survey, but only for pointed observations, thus leading to a decrease in the numbers of detections. Secondly, the source extension has not been taken into account. This effect can usually be estimated by degrading the sensitivity of the instrument linearly by the source apparent size when it is larger than the PSF of the instrument~\citep{dubus2013}.
 
Another effect which will be important is source confusion. The angular extension of the SNRs, the unknown level of the diffusion emission, and the high numbers of the sources leading to overlapping between sources may be a significant issue for identification. This effect is expected to be especially relevant in the inner regions of the Galaxy and for the fainter objects~\citep{hewitt2015}. According to the latest gamma--ray surveys, the most numerous Galactic sources are Pulsar Wind Nebulae (PWNe) and the overlapping of sources is expected~\citep{donath2016}. 

We propose to estimate the effect of the overlapping by the following method. Let us consider overlaps between SNRs and PWNe. The effect of the diffuse emission is not taken into account, since rough extrapolation to the TeV of the diffuse emission measured in the GeV range~\citep{ackermann2012} suggests that the detection of SNRs will be significantly affected for sources with apparent size of the order of a few 0.1$^{\circ}$. 
    For each realization of the time and location of supernovae in the Galaxy that we simulate (as described in Sec.~\ref{sec:montecarlo}), we simulate the locations and sizes of PWNe in the region of interest, the inner part of the GPS of CTA ($| l  | < 60^{\circ}$, $| b | < 2^{\circ}$) and we count the number of overlaps between the simulated PWNe and simulated SNRs.
    The simulation of the PWNe is done as follows:~\citet{dubus2013} estimated that about 300 to 600 PWNe should be detectable depending on the final performances of the telescope. Following the Galactic source distribution model of~\citet{renaud2011}, the number of sources per square degree along each line--of--sight within $| l  | < 60^{\circ}$, $| b | < 2^{\circ}$ can be estimated. First order (i.e. neglecting the local variations at the spiral arm tangents), the resulting Galactic distribution of PWNe is well fitted with a two--dimensional Gaussian at the Galactic center position, with a standard deviation of $\approx 40^{\circ}$ and $\approx 0.5^{\circ}$ in Galactic longitude $l$ and latitude $b$ respectively, and a maximum value in the Galactic center region of $\approx 4 (N_{\rm PWNe}/500)$ sources per square degree (where $N_{\rm PWNe}$ is the total number of PWNe in the region of interest). 
In this case, CTA could detect up to $\approx 200$ sources in the central regions of the Galaxy $| l  | < 30^{\circ}$, $| b | < 0.5^{\circ}$ and $\approx 500$ sources in the region $| l  | < 60^{\circ}$, $| b | < 2^{\circ}$. The population of PWNe is expected to be middle--aged so that their extension reaches scales of the order of $\approx 0.1^{\circ}- 0.3^{\circ}$ at a few kpc. 
We study the impact of the average size of the PWNe on the number of confused SNRs (overlaps between SNRs and PWNe, divided by the number of SNRs). For an average size of $0.1^{\circ}$, the fraction of confused SNRs is negligible even considering 600 PWNe. For an average PWNe size reaching $0.3^{\circ}$, the fraction of confused SNRs is of $\approx 0.1$ and $\approx 0.2$ in the cases 300 and 600 PWNe, therefore significantly affecting the detected SNRs.  
In this study we also take into account the source confusion due to the superposition of SNRs with other SNRs. We estimate this effect in the case where it is expected to be the most relevant (Model M1), which is the case where the number of detection is the largest. We found that the fraction of SNR--SNR superposition is leading to a confusion fraction smaller than $\approx 0.02$. The confusion is therefore expected to be dominated by the superposition with PWNe. 
In addition, during the H.E.S.S survey, the number of young SNRs detected was found roughly equal the number of mixed--morphology SNRs, from which the gamma emission is ambiguously due to the shell or the PWN~\citep{donath2016}. This suggests a loss in useful detection of a factor $\approx 2$. In the case of CTA, this factor could be reduced, given the improved angular resolution compared to H.E.S.S. but will affect the detected population, favoring smaller (i.e. more distant, younger) sources. 

Other very--high--energy gamma--ray sources such as stellar clusters or star forming regions are not taken into account in this work. The confusion with these sources could be estimated in an analogous way, by simulating the number and apparent sizes of these regions, and counting the number of overlappings. By extrapolating the results of the H.E.S.S. GPS, one can argue that the confusion due to these regions is expected to be smaller than the one due to the PWNe population.

\begin{figure}
\includegraphics[width=.5\textwidth]{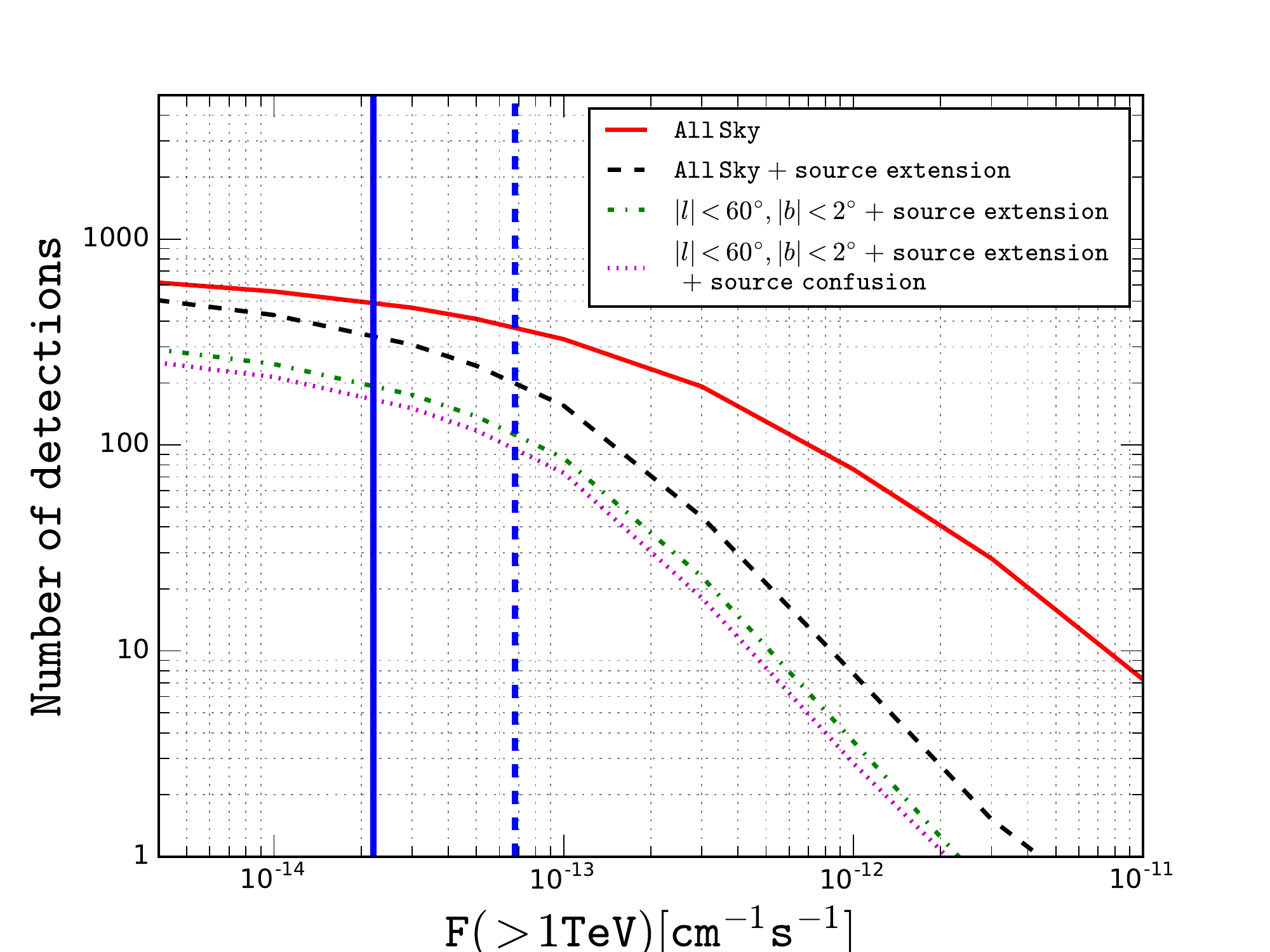}
\caption{SNRs with integral gamma--ray flux above F($>1$ TeV) in the case of model M1. The red (solid) curve corresponds the entire Galaxy, the black (dashed) line corresponds to the entire Galaxy and takes into account the effect of the extension of sources. The green (dot--dashed) line corresponds to the inner part of the Galaxy ($| l  | < 60^{\circ}$, $| b | < 2^{\circ}$) and takes into account the effect of the extension of sources. The magenta (pointed) line corresponds to the same situation that the previous one, but adding the effect of source confusion described in Sec.~\ref{sec:CTA}.
The blue solid and dot--dashed vertical lines correspond to a sensitivity of 1 mCrab and 3 mCrab, respectively.}
\label{fig:confusion}
\end{figure}

As an example of these effects and how they can affect the population of SNRs detected during the GPS, we represent in Fig.~\ref{fig:confusion} the number of SNRs with integral fluxes above  F($>1$ TeV) in the case of model M1, which is the model leading to the largest number of detections in our study. 
The red (solid) line corresponds to the case of M1, considering the entire Galaxy, which is the  situation described by the solid (red) line of Fig.~\ref{fig:LNLS1}. Taking into account the extension of sources (black dashed line), the number of detections is reduced by a factor~$\approx 1.4-1.8$ for sensitivity in 1-3 mCrab. At a sensitivity of 3 mCrab, the number of potentially detectable SNRs drops from $\approx 370$ to $\approx 200$.  The effect becomes negligible for fluxes around $\approx 10^{-15}$ cm$^{-1}$s$^{-2}$. We continue by adding the effect (green dot--dashed line) of reducing the survey to a portion of the Galactic plane $| l  | < 60^{\circ}$, $| b | < 2^{\circ}$, leading to a reduction of roughly factor~$\approx 1.7-1.9$ compared to the previous case. At a level of 3 mCrab, the number of potentially detectable SNRs from $\approx 200$ to $\approx 120$. Finally, the effects of source confusion are added (magenta pointed line). At the level of the sensitivity of CTA, the detection is affected by a factor of~$\approx 1.2$, which corresponds at 3 mCrab, to a number of $\approx 100$ SNRs. 
We remark that at a level of $10^{-11}$ cm$^{-2}$s$^{-1}$, taking into account the reduced survey, the source extension and the source confusion, the mean number of detection is $0.2^{+2}_{-0.2}$. In this region, one SNR, RXJ 1713.7--3946~\citep{RXJ2016}, has been detected with such flux. Our results are therefore still compatible with this observation, but illustrates than the computing of the different confusion and extension effects is a first order approximation.

These numbers illustrate how the different effects presented above can be taken into account in our model, to simulate the situation corresponding to the performed GPS survey of CTA.

Two strategies have been proposed for the Galactic Plane Survey of CTA. A survey of the entire Galactic Plane, where the sensitivity above 1 TeV is typically expected to be $\approx$ 3 mCrab, and an extensive deeper survey in the central region ($| l  | < 60^{\circ}$, $| b | < 2^{\circ}$) , where the sensitivity could reach the order of $\approx$ 1 mCrab. 
We consider these two possible strategies and provide the typical description of the obtained population, considering the three models described previously. Our results are presented in Table~\ref{tab:numbersCTA}.
In this example, the extension of the sources has been taken into account. The integral fluxes have been degraded linearly by a factor of~$\vartheta_s/\vartheta_{\rm PSF}$, where $\vartheta_s$ is the apparent size of the source and $\vartheta_{\rm PSF} \approx 0.05^{\circ}$ is the angular resolution of CTA~\citep[see e.g.][]{felixreview,aharonian1997,HAWC2008}. 
Naturally, as we go from model M1 to M3, thus from hard to steep spectra and from high to low values of the electron--to--proton ratio, the average number of detection of sources decreases. 
The number of detections can already be compared to the known SNR shells, in the whole Galaxy or in the region of interest. The known number of shells detected in gamma rays in the entire Galaxy is $\gtrsim 13$~\citep[see e.g.][describing the TeVCat]{TeVCat}, although one has to be careful because this number does not account for the circumstances of discovery of these shells, which might have been detected because of their interactions with other objects or extensive targeted observations.  This number can be compared with the number of potential detection for Model M3, 8, suggesting indeed that the parameters of M3 are in tension with observation. This is consistent with the fact that several of the observed Galactic SNRs are thought to  have their gamma--ray emission dominated by leptonic mechanisms, and therefore a model accounting only for hadronic sources does not seem to agree with observations~\citep{aharonian2013}.
From Table~\ref{tab:numbersCTA} we remark that the increased sensitivity in the inner part of the GPS results logically in the detection of more distant and older SNRs. The fraction of point--like sources varies in the different models and observation strategies in the range 15$\%$-- 50$\%$. The models leading to the greater number of detections accounting for the detection of the most extended sources.  

The simulated populations of SNRs can be compared to the ones obtained for simulations of the H.E.S.S. GPS. M1 and M2 correspond respectively to model M6 and M5 in~\cite{cristofari1}. We remark that the median distances of the detected SNRs in the CTA GPS simulations in the different models are larger by a factor $\approx 1.3 -1.6$, naturally illustrating that the improved sensitivity helps to detect SNRs located further away. 
In the case of the inner GPS of CTA the sensitivity above 1 TeV is typically improved by a factor of $\approx 5$ compared to H.E.S.S. GPS (from 15 mCrab to 3 mCrab), and for surveys of comparable extension ($| l  | <40^{\circ}$, $| b | < 3^{\circ}$ in the former H.E.S.S. GPS, and $| l  | <60^{\circ}$, $| b | < 2^{\circ}$ for the inner CTA GPS). If we consider, as an approximation, SNRs to be uniformly distributed in a flat disc, the improvement in sensitivity would lead to SNRs visible up to a distance $\approx 5^{1/2}$. In our case the improvement in the mean distance is found in the range $\approx 1.3 - 1.6$, accounting for the fact that SNRs are not uniformly distributed, and that the survey regions are not exactly equivalent.

The median ages are found larger by a factor $\approx 1.5$, accounting for the detection of older SNRs. The improved sensitivity of CTA leads to a greater number of potentially detectable SNRs and thus the possibility to detect fainter and older objects. At the level of 1 mCrab and in the hypothetical most optimistic case (M1), about $N \approx 500$ SNRs are potentially detectable. If we consider a rate of $\nu_{\rm SN}$ 3 SN/century and that these SNRs are uniformly aged, the median age of the population is $\frac{1}{2}\times (N / \nu_{\rm SN}) \approx 8$ kyr. 
This is without taking into account the source extension and the source confusion described above. Both of these effects  select less extended and thus younger SNRs, so that the median age found for M1 is $\approx 5$ kyr, as presented in Tab~\ref{tab:numbersCTA}.

The median size of the resolved sources is comparable to the one found in the H.E.S.S. GPS study, and we found that the fraction of resolved sources is a factor of $\approx 2$ greater in the CTA GPS. This effect can be explained by the improved angular resolution of CTA, here typically $\approx 0.05^{\circ}$ at 1 TeV, compared to $\approx 0.1^{\circ}$ for H.E.S.S. 
Finally the fraction of hadronic, defined as the contribution from hadronic interactions to the gamma--ray luminosity relative to the total gamma--ray luminosity at 1 TeV, is a factor of $\approx 2$ smaller than in the H.E.S.S. study. This suggests that the improved performance of CTA will lead to a greater number of SNRs with gamma--ray emission dominated by leptonic mechanisms.

\begin{table*}
\centering
\begin{tabular}{l c c c c c c}
\hline 
Survey: & \multicolumn{3}{|c|}{full GPS}& \multicolumn{3}{|c|}{inner GPS}  \\
\hline 
Model: &  M1  & M2 & M3  &  M1  & M2 & M3 \\
\hline 
Median number of detections: & $200$ &  $21$ &  $7$ &$190$ &  $46$ &  $18$\\
Median distance [kpc]: & 8.5 &4.5  & 5.1 & 10 &8.1 & 7.2 \\
Median age [kyr]: &5.0  & 1.7 & 0.61 &6.7  & 4.1 & 4.4 \\
Median apparent size [$^{\circ}$]*: & 0.21 & 0.20 & 0.06 & 0.18 & 0.20  &0.09 \\
Fraction of point sources:  & 0.17 & 0.26&  0.48 &  0.15&  0.24  &  0.39 \\
Fraction of hadronic sources: &  0.3 &  0.4 & 1 & 0.3 &0.4  & 1\\
\hline
\scriptsize{* Extended sources only (i.e. size larger than 0.05$^{\circ}$).}
\end{tabular}
\caption{Characteristics of the simulated SNR population detected in the Galactic plane survey of CTA  for the different models listed in Table~\ref{tab:modelsCTA}. The source extension has been taken into account by degrading  the integral fluxes by a factor of  $\vartheta_s/\vartheta_{\rm PSF}$, where $\vartheta_s$ is the source apparent size and $\vartheta_{\rm PSF} \approx 0.05^{\circ}$ is the angular resolution of CTA. The \textit{full GPS} and \textit{inner GPS} survey correspond to a sensitivity at 1 TeV of 3 mCrab and 1 mCrab respectively.}
\label{tab:numbersCTA}
\end{table*}

A discussion of the influence of several parameters used in this model on our results, such as the rate of Galactic supernova explosion, the structure of the magnetic field, or the maximum energy of accelerated particles at the shock can be found in~C13. We remark that adopting other realistic descriptions for these parameters tend to decrease the numbers of expected detections, highlighting the fact that results presented in this Section should be seen as upper limits. The most influential parameter in our study appears to be $p_{\rm max}$. Adopting a different more pessimistic (but still plausible) description of $p_{\rm max}$ leads to numbers of potentially detectable SNRs reduced by a factor less than $\lesssim 2$.

 \section{Conclusions}
 This paper presents estimates of the SNR population that CTA can expect to detect in the TeV domain. The population study of SNRs that we presented, based on Monte Carlo simulation, aims at being confronted with the results of the CTA Galactic plane survey. This future confrontation will help test again the SNR paradigm for the origin of CRs, and constrain the parameters governing particle acceleration at SNR shocks. 
 
 In the presented work, we have assumed that SNRs are the main sources of CRs, and estimated the typical acceleration efficiency per SNR. A Monte Carlo approach was used to simulate the time and explosion of supernovae in the Galaxy, and to estimate the number of SNRs expected to be detected by CTA. This work had been previously tested on the available data in the TeV domain, and the authors believe that this approach is therefore relevant in the TeV and multi--TeV domain for CTA. 
We have presented the typical populations expected for different extreme values of the electron--to--proton ratio $K_{\rm ep}$ and the slope of accelerated particles at SNR shocks $\alpha$, and found that in the TeV range, these extreme situations lead to significantly different results. The numbers of SNR detections presented for the observation of the entire Galaxy at the level of $\approx $ 1 mCrab are $200^{+20}_{-20}$, $21^{+5}_{-5}$ and  $8^{+3}_{-3}$ for Model M1, M2 and M3 respectively. 
These results suggest that a  confrontation with the observation of the CTA Galactic plane survey could directly give powerful insight on the values of the parameters governing particle acceleration. The numbers of SNR detections in the GPS will be a first direct indicator, and an extensive description of the characteristics of the population could provide arguments in favor of a set of parameters. 
More models could obviously be considered to constrain more finely the acceleration parameters. 

Moreover, it is remarkable that SNRs have been historically detected in the radio wavelength range. \citet{green2014,green2015} reports a catalog of $\lesssim 300$ SNRs, and our results suggests that CTA alone could detect a number of SNRs of the same order, therefore becoming a very efficient tool for the detection of SNRs. 

These models and values that we have chosen to present correspond to an ideal and optimistic description of the particle acceleration at SNR shocks. 
Several effects presented in Sec.~\ref{sec:CTA} are expected to affect these results, such as the extension of the sources, a reduction of the Galactic plane survey, and the problem of the identification of new SNRs as such. We have described in Sec.~\ref{sec:CTA} how these effects can be quantitively taken into account. It will therefore be crucial to implement them before confronting the actual CTA observations. 
We also mention that theoretical advances or observation that would lead to a better description of any of the parameters used in our model, could be implemented in our approach. 
%A fiercer constrain on any of the open parameters, confronted with observations,  would then directly lead to reduce the acceptable range of values of the remaining parameters. 

The work presented in this paper describes a method which could also be used to investigate the performance of other instruments operating in the TeV range and in the multi--TeV range, such as the HAWC observatory~\citep{HAWC2016}, LHAASO~\citep{LHAASO2016}, or HiSCORE~\citep{HiSCORE2014}, therefore providing more tools to improve our understanding of the origin of Galactic cosmic rays. This work will be carried in a forthcoming paper.

In this paper, we have chosen not to discuss the problem of PeVatrons, though it is indeed a crucial question that instruments operating in the TeV and multi--TeV range will help to address. In the SNR hypothesis for the origin of Galactic CRs, it is often implied that the acceleration of particles up to PeV energies has to be provided by SNR shocks. Observations in the multi--TeV range, where efficient gamma ray production from leptonic interactions is not possible due to the Klein--Nishina cut-off, could help solve this problem, and therefore provide an unequivocal proof that SNRs are PeVatrons. 
The clear identification of SNR PeVatrons as such has not been provided by current instruments operating in the TeV range, but the new VHE instruments will provide increased chances of  detecting such objects. 
The model presented in this article can be used to investigate the problem of the detection of PeVatrons, and will be presented in a forthcoming paper.

\section*{Acknowledgments}

We thank M. Bottcher, S. Covino, L. Drury, E. M. De Gouveia Dal Pino, E. De Ona Wilhelmi, J.-P. Ernenwein, J. Holder, A. Marcowith, R. Mukherjee, J. Mart\'{i}, T. Montmerle, M. Renaud and R. Zanin for helpful discussions. 
SG acknowledges support from the Programme National Hautes Energies (CNRS). SG and RT acknowledge support from the Observatoire de Paris (Action F\'ed\'eratrice Preparation à CTA). PC acknowledges support from the Columbia University Frontiers of Science fellowship.

\end{document}